\documentclass[]{aa}
\usepackage[switch]{lineno}
\usepackage{graphicx}
\usepackage{arydshln}
\usepackage{subcaption}
\usepackage{hyperref}
\hypersetup{colorlinks=true,linkcolor=blue,filecolor=blue,urlcolor=blue,citecolor=blue}
\usepackage{appendix}
\usepackage{comment}
\usepackage{anyfontsize}
\usepackage{orcidlink}
\usepackage{multirow}

\newcommand{\gamsh}{\gamma_\mathrm{sh}}
\newcommand{\gamz}{\gamma_\mathrm{0}}

\newcommand{\mpr}{m_\mathrm{p}}
\newcommand{\me}{m_\mathrm{e}}
\newcommand{\tp}{T_\mathrm{p}}
\newcommand{\te}{T_\mathrm{e}}
\newcommand{\gad}{\Gamma_\mathrm{ad}}

\newcommand{\lsi}{\lambda_\mathrm{si}}

\usepackage{txfonts}
\makeatletter
\renewcommand*\aa@pageof{, page \thepage{} of \pageref*{LastPage}}
\makeatother

\begin{document}

   \title{Radiative signatures of electron-ion shocks in BL Lac type objects}
   \titlerunning{Radiative signatures of electron-ion shocks}

   \author{A. Arbet-Engels\inst{1}\orcidlink{0000-0001-9076-9582}, 
           A. Bohdan\inst{2}\orcidlink{0000-0002-5680-0766},
           F. Rieger\inst{2}\orcidlink{0000-0003-1334-2993},
           D. Paneque\inst{1}\orcidlink{0000-0002-2830-0502}, \and F. Jenko\inst{2}\orcidlink{0000-0001-6686-1469}}

   \institute{Max-Planck-Institut für Physik, Boltzmannstr. 8, DE-85748 Garching, Germany\\
              \email{aarbet@mpp.mpg.de}
        \and
              Max-Planck-Institut für Plasmaphysik, Boltzmannstr. 2, DE-85748 Garching, Germany\\
              \email{artem.bohdan@ipp.mpg.de, frank.rieger@ipp.mpg.de}
             }

   \date{Received ...; accepted ...}

   \date{Received XX XX, 2025; accepted XX XX, 2025}

 
  \abstract
   {}
   {Plasma shock waves stand out as one of the most promising sites of efficient particle acceleration in extragalactic 
   jets. In electron-ion plasma shocks, electrons can be heated up to large Lorentz factors, making 
   them an attractive scenario to explain the high minimum electron Lorentz factors regularly needed to describe 
   the emission of BL Lac type objects. Still, the (relativistic) thermal electron component is commonly neglected 
   when modelling the observations, although it holds key informations on the shock properties.}
   {Considering a shock acceleration scenario, we model the broadband emission of the archetypal high synchrotron 
   peaked (HSP) blazar Markarian~421 employing particle distributions that include a thermal (relativistic) Maxwellian 
   component at low energies followed by a nonthermal power-law, as motivated by particle-in-cell simulations. The observations, 
   in particular in the optical/UV and MeV-GeV bands, efficiently restrict the nonthermal emission from the Maxwellian electrons, 
   which we use to derive constraints on the basic properties, such as the fraction $\epsilon_e$ of the total shock energy stored
   in the nonthermal electrons.}
  {The best-fit model yields a nonthermal electron power-law with an index of $\sim 2.4$, close to predictions from shock 
  acceleration. Successful fits are obtained when the ratio between the Lorentz factor at which the nonthermal distribution begins 
  ($\gamma_{\rm nth}$) and the dimensionless electron temperature ($\theta$) satisfies $\gamma_{\rm nth}/\theta \lesssim 8$. 
  Since $\gamma_{\rm nth}/\theta$ controls $\epsilon_e$, the latter limit implies that at least $\epsilon_e \approx 10\%$ of 
  the shock energy is transferred to the nonthermal electrons. These results are almost insensitive to the shock velocity 
  $\gamma_{\rm sh}$, but radio observations indicate $\gamma_{\rm sh} \gtrsim 5$ since for lower shock velocities the fluxes 
  in the millimeter band are overproduced by the Maxwellian electrons. Hence, if shocks drive the particle 
  energisation, our findings indicate that they operate in the mildly to fully relativistic regime with efficient electron acceleration. This paper sets the grounds for future works, 
  in which we will investigate with plasma simulations if, and under which conditions, the findings presented here can be reproduced.}
  {}
   \keywords{acceleration of particles -- shock waves -- radiation mechanisms:  nonthermal -- BL Lacertae objects:  individual (Markarian 421) -- galaxies:  jets}

   \maketitle
%


\section{Introduction} \label{sec:intro}

Plasma shock waves are ubiquitous phenomena in galactic sources of high-energy radiation. They efficiently convert 
kinetic energy of the plasma flow into the acceleration of particles up to relativistic speeds \citep[e.g.,][]{fermi1949,
bell1978,drury1983,blandford1987}. In extragalactic jets launched by active galactic nuclei (AGNs), mildly and/or fully 
relativistic shocks are also believed to play a central role in the energisation of charged particles up to $\sim$\,TeV 
energies \citep{blandford1979,marscher_gear1985,kirk1998}. Some of these regions may be associated with superluminal 
features observed at radio frequencies \citep[e.g.,][]{zensus1997}. Additional observational evidence was recently brought 
by the measurement of the polarisation up to the X-rays in the synchrotron spectrum of BL Lac type objects. Both in 
quiescent and flaring states, the X-ray polarisation angle aligns on average well with the angle measured at lower 
frequencies as well as with the jet axis \citep[e.g.,][]{liodakis2022,magic_ixpe_mrk421flare2023, mrk421_2022, 
digesu_acceleration2022}. 
This behaviour is expected if a (self-generated) magnetic field component orthogonal to the jet axis develops at the 
shock front, as revealed by particle-in-cell (PIC) simulations \citep{crumley2019, vanthiegem2020}.\par 

In the specific case of electron-ion plasma shocks, most of the available shock energy is contained in the ions. Part of 
the shock energy heats the electrons to a fraction of the equipartition, leading to the development of thermal (relativistic) 
Maxwellian distribution that can easily peak at Lorentz factors of $\gamma \sim 10^2-10^3$, even for mildly relativistic 
shocks. This motivates to consider electron-ion shocks as potential cause to explain the relatively high minimum 
electron Lorentz factors required to describe the spectral energy distribution (SED) of ``extreme TeV'' BL Lac objects \citep{kaufmann2011, costamante2018,biteau2020,zech_lemoine2021}.\par 

While PIC simulations of shocks indicate a thermal Maxwellian distribution in the low-energy part of the particle population, the modelling of the broadband emission from BL Lac objects (and blazars in general) is typically performed without considering this component \citep{boettcher2013, tavecchio2010}. The radiating particles are modelled with a power-law, broken power-law or log-parabola distributions with a hard cut-off at low energies. However, even if located at the lowest energies, the nonthermal radiation signature of the Maxwellian component might in principle be detectable with gamma-ray instruments. Recently, \cite{tavecchio2025} investigated the case of flat spectrum radio quasars (FSRQs), unveiling that a thermal particle population in electron-ion shocks may 
be constrained by MeV observatories.\par 

Constraining the (relativistic) thermal distribution is highly relevant to extract fundamental properties of the shock, such as the fraction of shock and thermal energy that is conveyed to the nonthermal  particles. Those properties can be subsequently tested with PIC simulations in order to further constrain shock parameters. Here, we study the case of BL Lac type objects, and in particular high synchrotron peaked \citep[HSPs;][]{abdo_hsp_2010} blazars. For HSPs, the nonthermal emission from the Maxwellian electrons in mildly and full relativistic electron-ion shocks is expected in the millimeter-to-infrared range as well as in the MeV-GeV gamma-ray band \citep[assuming leptonic models; ][]{mastichiadis_kirk_1997, krawczynski2004}. Those energies, in particular the latter, are well covered by currently operating instruments. Exploiting a well sampled SED from \citet{abdo:2011} of the HSP Markarian~421 (hereafter Mrk~421), we model the nonthermal emission with a PIC motivated electron distribution in order set limits on basic properties of the Maxwellian distribution. This work is part of a broader effort, where in a second step we plan to verify with dedicated PIC simulations if, and under which conditions of electron-ion shocks, the findings of this work can be matched.\par

The paper is structured as follows. Sect.~\ref{sec:model_description} and Sect.~\ref{sec:application} describe the model and its application to Mrk~421. The results are shown in Sect.~\ref{sec:results}, and in Sect.~\ref{sec:shock_simul} we discuss their implications for the shock parameters. Finally, the conclusions are drawn in Sect.~\ref{sec:discussion}.

\section{Model setup} \label{sec:model_description}

For simplicity and clarity, we consider an electron-proton plasma shock with an equal number density between the two particle species, $\eta := n_{\rm e} / n_{\rm p} = 1$. Beyond the shock, the energetic particles are injected in the downstream region, where they emit nonthermal emission. As demonstrated by PIC simulations \citep[e.g.,][]{sironi_spitkovsky_2011, crumley2019}, the low-energy part of the particle population is heated and forms a thermal component following a (relativistic) Maxwellian distribution. At higher energies, $\gamma > \gamma_{\rm nth}$, a nonthermal power-law component develops as a result of first-order Fermi type shock acceleration. 
Such distribution can be parametrised as described in \cite{giannios_spitkovsky_2009}:\par 
\begin{equation}
\label{eq:electron_distribution_gia_spit}
    \frac{dn_{\rm e}}{d\gamma}(\gamma)\propto \begin{cases}
    \mathcal{N}(\gamma,\theta), \; 1<\gamma<\gamma_{\rm nth}\\
    \mathcal{N}(\gamma_{\rm nth},\theta) \, \left(\frac{\gamma}{\gamma_{\rm nth}} \right)^{-p_1}, \; \gamma_{\rm nth} < \gamma ,\\
    \end{cases}
\end{equation}
where $\mathcal{N}(\gamma,\theta) = \frac{\gamma^2}{2\theta^3} \, e^{-\gamma/\theta}$ describes the Maxwellian 
thermal component, $\theta\equiv k T_{\rm e}/  m_{\rm e} c^2$ is the dimensionless particle temperature, $p_1$ 
is the power-law index of accelerated particles and $\gamma_{\rm nth}$ designates the Lorentz factor at which 
the nonthermal particle distribution starts. The above parametrisation for the thermal component ($\mathcal{N}(\gamma,\theta)$) is a generic functional form of a (relativistic) Maxwellian distribution valid in the limit where $\theta>>1$. As demonstrated in the following sections, this is expected even for mildly relativistic electron-ion shocks, and the fit results presented later further constrain $\theta>>1$. The adopted parametrisation is thus well-justified.\par 

In the downstream region, we calculate the resulting nonthermal radiation in a framework of a purely leptonic 
model using the same code as used in \cite{mrk421_2017}. The broadband spectrum is ascribed to synchrotron 
self-Compton (SSC) emission, which is the simplest scenario that satisfactorily describes SEDs of BL Lac, 
and HSPs in particular \citep[see e.g.,][]{mastichiadis_kirk_1997, Tavecchio_Constraints, krawczynski2004}. The code incorporates synchrotron self-absorption effects at radio frequencies using the equations in Appendix~C of \citet{zacharias2013}.
We further assume that particles escape the emitting region, approximated as a sphere, on timescales of 
$t_{\rm esc}=R/c$, where $R$ is the radius of the region in the comoving frame. The emission is calculated 
by assuming that the particle distribution reached its equilibrium state, taking into account particle escape 
and synchrotron radiation as the dominant cooling effect. In all the cases investigated in this work, synchrotron 
cooling becomes effective for electrons with Lorentz factor $\gamma \gtrsim 10^5$, i.e., deep in the nonthermal 
part of the population. For $\gamma \lesssim 10^5$, particles escape the downstream region before radiating a 
significant fraction of their energy. The equilibrium state of the particle distribution therefore has a spectral 
break of $\Delta p = 1$ in the power-law where the electron cooling timescale equals the escape time, i.e. at 
$\gamma_{\rm br}=\frac{6 \pi m_{\rm e} c^2}{\sigma_T B^2 R}$, where $\sigma_T$ is the Thomson cross section 
and $B$ the comoving magnetic field \citep{kirk1998}. Consequently, the electron distribution described by Eq.~\ref{eq:electron_distribution_gia_spit} is modified as follows:
\begin{equation}
\label{eq:electron_distribution}
    \frac{dn_{\rm e}}{d\gamma}(\gamma)\propto \begin{cases}
    \mathcal{N}(\gamma,\theta), \; 1<\gamma<\gamma_{\rm nth}\\
    \mathcal{N}(\gamma_{\rm nth},\theta) \, \left(\frac{\gamma}{\gamma_{\rm nth}} \right)^{-p_1}, \; \gamma_{\rm nth} < \gamma < \gamma_{\rm br},\\
    \mathcal{N}(\gamma_{\rm nth},\theta) \, \left(\frac{\gamma_{\rm br}}{\gamma_{\rm nth}} \right) \, e^{\gamma_{\rm br}/\gamma_{\rm cut}} \, \left(\frac{\gamma}{\gamma_{\rm nth}} \right)^{-p_1-1} \, e^{-\gamma/\gamma_{\rm cut}}, \; \gamma > \gamma_{\rm br},
    \end{cases}
\end{equation}
where $\gamma_{\rm cut}$ is the cut-off energy. If the electron temperature ($\theta$) is sufficiently high, 
this distribution effectively mimics one with a high $\gamma_{\rm min}$ (cf. Fig.~\ref{eed_sample_distr}).
We employ this parametrisation for the SED modelling presented below.

\begin{figure}[h!]
\begin{center}
 \includegraphics[width=\columnwidth]{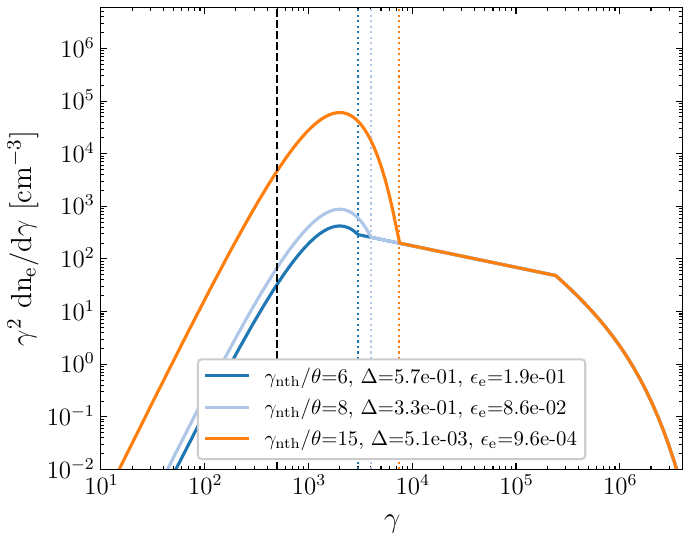}
\end{center}
 \caption{Sample of electron distributions using Eq.~\ref{eq:electron_distribution} for different $\gamma_{\rm nth} / \theta$ values ranging from 6 to 15. The black dashed line represents $\theta$, which is fixed to 500. The colored dotted lines give the location of $\gamma_{\rm nth}$ where the transition from the Maxwellian to the nonthermal component occurs. Here, we arbitrarily set $\gamma_{\rm br}=2\times10^5$ and  $\gamma_{\rm cut}=7\times10^5$ and $p_1=2.4$.}
 \label{eed_sample_distr}
\end{figure}

Two additional properties are conventionally used to characterise the particle distribution resulting from an electron-proton shock: the fraction $\Delta$ of the total electron energy residing in the nonthermal component, and the fraction $\epsilon_e$ of the total shock energy carried by the nonthermal electrons. We define $\Delta$ as in \cite{giannios_spitkovsky_2009}:

\begin{equation}
\label{eq:Delta_def}
    \Delta =  \frac{\int_{\gamma_{\rm nth}}^{\infty} \gamma \, \frac{dn_{\rm e}}{d\gamma}\, d\gamma} {\int_{1}^{\infty} \gamma \, \frac{dn_{\rm e}}{d\gamma}\, d\gamma } \ ,
\end{equation}
while $\epsilon_e$ is given by:

\begin{equation}
\label{eq:epsilon_e}
    \epsilon_e = \frac{m_{\rm e} c^2 \int_{\gamma_{\rm nth}}^{\infty} \gamma \, \frac{dn_{\rm e}}{d\gamma} d\gamma} {n_{\rm p} m_{\rm p} c^2 (\gamma_0 -1)} \ ,
\end{equation}
where $n_{\rm p}$ is the proton number density and $\gamma_0$ is the bulk Lorentz factor of the upstream plasma (jet plasma before the shock) as observed in the downstream frame (the emitting region). $\gamma_0$ depends on several factors such as $\theta$ and the temperature ratio between the protons and electrons, which we assume to be $T_e/T_p=0.5$ for the rest of this work \citep[that is compatible with PIC simulations of relativistic electron-ion shocks, see][]{sironi_spitkovsky_2011}. The corresponding functional form of $\gamma_0$ is derived in Sect.~\ref{sec:shock_simul}. Following \citet{tavecchio2025}, the above expression can be simplified by introducing $\bar{\gamma}$, the average electron Lorentz factor, and $\eta$:

\begin{equation}
    \epsilon_e = \frac{m_{\rm e}}{m_{\rm p}} \eta \frac{\bar{\gamma} \; \Delta}{\gamma_0 -1}, \quad \mathrm{where} \quad \bar{\gamma} = \frac{\int_{1}^{\infty} \gamma \, \frac{dn_{\rm e}}{d\gamma}\, d\gamma} {\int_{1}^{\infty} \frac{dn_{\rm e}}{d\gamma}\, d\gamma } \ .
\end{equation}
As already mentioned, in what follows we use $\eta =1$.\par 

It is important to note that the ratio $\gamma_{\rm nth} / \theta$ directly dictates $\Delta$, and therefore also $\epsilon_e$. A higher $\gamma_{\rm nth} / \theta$ value leads to a more prominent Maxwellian component with respect to the nonthermal part, at the cost of reducing $\Delta$ and $\epsilon_{\rm e}$. Fig.~\ref{eed_sample_distr} shows the electron distribution using Eq.~\ref{eq:electron_distribution} for several $\gamma_{\rm nth} / \theta$ values from 6 to 15. The corresponding values of $\Delta$ and $\epsilon_e$ are given in the legend.\par

\section{Application to the TeV BL Lac object Mrk~421} \label{sec:application}

We apply our model to the archetypal TeV BL Lac object Mrk~421 ($z=0.031$). Due to its brightness and proximity, Mrk~421 is one of the blazars whose broadband SED shape can be determined with exceptional precision even when the source shows very low activity \citep[for recent multiwavelength campaigns, see e.g., ][]{2016ApJ...819..156B,mrk421_2017, mrk421_2022}. Using a precise broadband SED of Mrk~421, we intend to constrain the contribution of a Maxwellian component to set limits on the shock parameters, focusing on $\Delta$ and $\epsilon_{\rm e}$.\par   

In the case of mildly relativistic shocks with $\gamma_{\rm sh} \sim (2-5)$, equivalent to $\theta \sim (10^2- 5\times10^2)$ (see Sect.~\ref{sec:shock_simul} and Eqs.~\ref{eq:gamma0totheta} and \ref{eq:gammashock}), the maximum
of the Maxwellian distribution will be located at $\gamma^{\rm peak}_{\rm th, e} \approx 3\theta \approx 10^2-10^3$. 
If we associate this peak with an effective minimum Lorentz factor, high values of $\gamma_{\rm min} 
\sim 10^3$ could be accommodated.  
The synchrotron emission of the thermal electrons is then expected to peak at an (observed) frequency of
\begin{equation}
\nu_{\rm syn} = 3.7 \times 10^6 \; \left(\gamma^{\rm peak}_{\rm th, e}\right)^2 \; B \delta \;\; \mathrm{Hz} \ .
\end{equation}

where $\delta$ is the Doppler factor of the emitting region. Considering typical values of magnetic field and Doppler factor inferred from blazar SED modelling
\citep[$B\sim0.1$\,G and $\delta\sim20$; ][]{tavecchio2010, abdo:2011}, $\nu_{\rm syn}~\sim~ 10^{11}-10^{14}$\,Hz,
i.e., in the millimeter to infrared band. As for their inverse-Compton contribution, the emitted frequency in 
one-zone SSC models is approximately (assuming the Thomson regime, as is typically the case for Lorentz factors 
of $\gamma^{\rm peak}_{\rm th, e} \approx 10^2-10^3$):
\begin{equation}
\nu_{\rm IC} = \frac{4}{3} \; \left(\gamma^{\rm peak}_{\rm th, e}\right)^2 \; \nu_{\rm syn, peak} \; \; \mathrm{Hz} \ ,
\end{equation}
where $\nu_{\rm syn, peak}$ is the peak frequency of the synchrotron component in the SED \citep{Tavecchio_Constraints}. In the specific case of the HSP Mrk~421, this SED peak frequency
$\nu_{\rm syn, peak}$ is around $10^{17}$\,Hz, implying a possible contribution of the Maxwellian 
component from $\sim10$\;MeV to $\sim10$\,GeV depending on the exact value of $\theta$ (and thus $\gamma^{\rm peak}_{\rm th, e}$). The MeV-GeV band is ideally covered by gamma-ray observatories such as \textit{Fermi}-LAT \citep{atwood2009, ackermann2012}, thus providing the opportunity to constrain the contribution of the Maxwellian hump. In passing, we note that since the radiative signature of Maxwellian electrons is expected at the rising edges of the two SED components, the hints of narrow features at the falling edges of the two SED components that were reported in the X-ray spectrum of Mrk~421 \citep[presented in][]{mrk421_2015_2016} and the TeV spectrum of the HSP Mrk~501 \citep[presented in][]{mrk501_2014_extremeflare} must have a different origin.\par

We use the broadband SED of Mrk~421 obtained during an average emission state published in \citet{abdo:2011}. 
The SED is averaged over a six-month period in 2009, during which moderate variability had been detected, suggesting the absence of significant change in the source properties and particle distribution. Hence, to model the particle distribution, it seems appropriate to employ a steady-state solution following Eq.~\ref{eq:electron_distribution}, which may result from an injection of particles at a constant rate in an emitting region whose physical properties stay roughly constant.\par

\begin{figure*}[h!]
    \centering
    \begin{subfigure}[b]{0.497\textwidth}
        \centering
        \includegraphics[width=\textwidth]{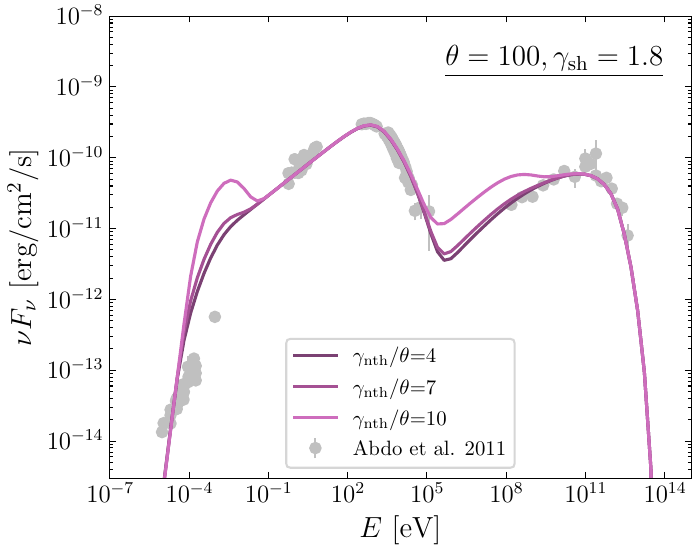}
    \end{subfigure}
    \begin{subfigure}[b]{0.497\textwidth}  
        \centering 
        \includegraphics[width=\textwidth]{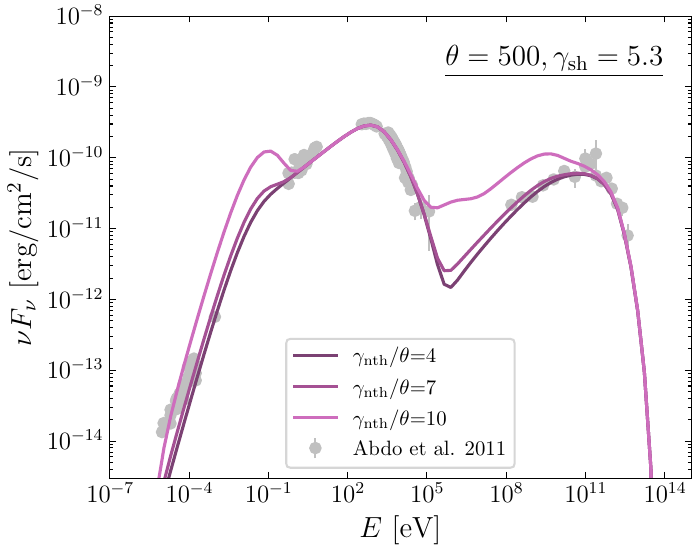}
    \end{subfigure}
    \caption{SED models for $\theta=100$ (left) and $\theta=500$ (right), corresponding to a shock velocity (in the upstream frame) of $\gamma_{\rm sh}=1.8$ and $\gamma_{\rm sh}=5.3$, respectively. We refer the reader to Sect.~\ref{sec:shock_simul} for more details on the connection between $\theta$ and $\gamma_{\rm sh}$. Solid lines, from light violet to dark violet, cover three different $\gamma_{\rm nth}/\theta$ ratios, 4, 7, and 10. The SED from \cite{abdo:2011} is plotted with grey markers.}
    \label{fig:SED_examples}
\end{figure*}

\begin{table}[h!]
\caption{\label{tab:ssc_param} SSC model parameters fixed throughout the work.}
\centering
\begin{tabular}{l c @{\hskip 0.3in} c}     
\hline\hline
Parameters   & Value  \\
\hline
\hline
$\delta$ & 20 \\
$R$ [$10^{16}$\,cm] & 4 \\
$B$ [$10^{-2}$\,G] & 5 \\
$p_1$ & 2.4\\
$\gamma_{\rm cut}$ & $7\times10^{5}$  \\
\hline 
\end{tabular}
\tablefoot{ We refer to Section~\ref{sec:application} for a detailed description of each parameter and the procedure adopted to determine them.}
\end{table}

We start by determining the parameters related to the source environment $\delta$, $R$, and $B$ as well as those of the electron distribution that can be derived independently of the Maxwellian component, that is, $p_1$ and $\gamma_{\rm cut}$. The peak luminosities and frequencies (that are unaffected by the Maxwellian component, unless in the most extreme cases) constrain $\delta$, $R$ and $B$ when considering a one-zone SSC model  \citep{Tavecchio_Constraints}. The $p_1$ and $\gamma_{\rm cut}$ parameters are constrained by the optical-to-X-ray part of the SED, in which the Maxwellian contribution is also negligible. $\gamma_{\rm br}$ is not a free parameter of the system and is given by the synchrotron cooling break (i.e., it is a function of $R$ and $B$). Prior to this initial fit, we set $\theta = \gamma_{\rm nth}$ such that the Maxwellian component is suppressed, and arbitrarily assume $\theta =10^3$, which satisfactorily describes the optical and MeV-GeV data without overproducing the radio fluxes. The SED is fitted by performing a scan over $\delta$ (from 10 to 30), and letting all other parameters free at each step. We then select the set of parameters with the lowest $\chi^2$ and for which $R < \delta \, c \, t_{\rm var}$, where $t_{\rm var}$ is the source variability timescale, which is $\approx 1$\,day \citep{abdo:2011}. The resulting set of parameters is given in Tab.~\ref{tab:ssc_param}. 
The inferred $\gamma_{\rm cut}$ is in principle compatible with expectations of first-order
Fermi acceleration at internal shocks \citep[e.g.,][]{rieger2007}.
The values reported here are very similar to those found by \cite{agnpy_paper} after 
applying different fitting procedures to the exact same SED of Mrk~421.\par

In a second step, we study the effect of the thermal component on the observed SED by scanning multiple $\theta$ and $\gamma_{\rm nth}$ values while fixing the other parameters to the ones of Tab.~\ref{tab:ssc_param} and keeping the synchrotron peak luminosity to a constant level by only modifying the electron density. Fig.~\ref{fig:SED_examples} show test cases for different $\gamma_{\rm nth}/\theta$ ratios, and $\theta=100$ (left panel) and $\theta=500$ (right panel). Such $\theta$ values correspond to a shock velocity (in the upstream frame) of $\gamma_{\rm sh}=1.8$ and $\gamma_{\rm sh}=5.3$, respectively, assuming a temperature ratio of $T_{\rm e}/T_{\rm p}=0.5$ between the electrons and protons. We refer the reader to Sect.~\ref{sec:shock_simul} for more details on the relation between $\theta$ and $\gamma_{\rm sh}$.\par 

The model starts to exceed the MeV-GeV part of the observed SED (covered by \textit{Fermi}-LAT) for $\gamma_{\rm nth}/\theta > 7$. The model also exceeds the optical data around $10^{-1}$\,eV for $\theta=500$ and $\gamma_{\rm nth}/\theta > 7$. These spectral features allow to set limits on $\gamma_{\rm nth}/\theta$, which controls the amount of energy transferred from the shock to the nonthermal electrons, hence constraining $\epsilon_e$ and $\Delta$. The prominence of the Maxwellian signature and its broad radiative pattern permit to set limits on $\gamma_{\rm nth}/\theta$ that are to a large extent independent on the exact value of the parameters such as $\delta$, $R$, $B$ (i.e., the parameters with the highest degree of degeneracy in one-zone models).\par 

We systematically explore the range of $\gamma_{\rm nth}/\theta$ allowed by the data by repeating the same procedure as for Fig.~\ref{fig:SED_examples}. We scan a wider range of $\theta$ and $\gamma_{\rm nth}/\theta$ ratios, and compute at each step the $\chi^2$ as a measure of the data/model mismatch. The $\chi^2$ is calculated above the radio band, i.e., beyond $10^{-1}$\,eV since due to synchrotron self-absorption it is known that one-zone SSC models fail to reproduce the radio spectra of blazars. The unsuitability of one-zone SSC models to simultaneously describe the VHE and radio spectra is thus independent of the used electron distribution. Hence, it is not meaningful to use a $\chi^2$ test to quantify the data/model mismatch in an energy range that is known apriori to be poorly reproduced by the model. Additionally, the radio band likely receives significant contribution from large scale structures in the jet that are broader and likely farther downstream than the compact region modelled here. In what follows, we will therefore consider the radio observations as upper limits to constrain the value of $\theta$ from below. The $\chi^2$ is calculated assuming a conservative 15\% systematic uncertainty on the fluxes for all measurements.

\begin{figure}[h!]
\begin{center}
 \includegraphics[width=\columnwidth]{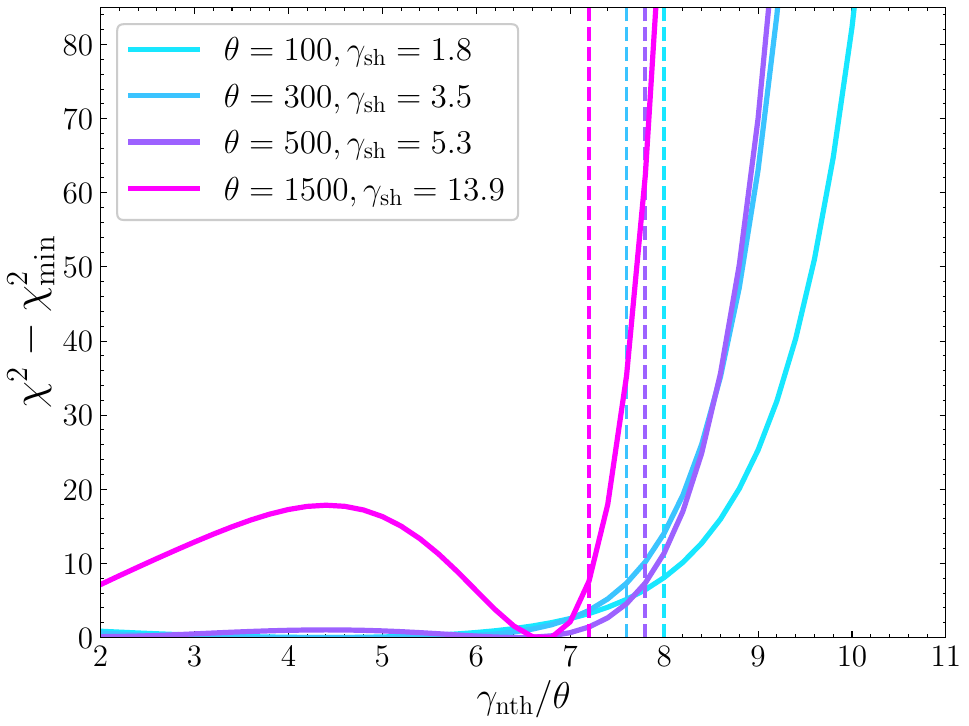}
\end{center}
 \caption{$\chi^2-\chi^2_{\rm min}$ as a function of $\gamma_{\rm nth}/\theta$, for $\theta ~=~[100, 300, 500, 1500]$, equivalent to a shock velocity of $\gamma_{\rm sh}~=~[1.8, 3.5, 5.3, 13.9]$. We refer the reader to Sect.~\ref{sec:shock_simul} for more details on the connection between $\theta$ and $\gamma_{\rm sh}$. $\chi^2_{\rm min}$ is the minimum $\chi^2$ for a given $\theta$. The vertical dashed lines mark the $\gamma_{\rm nth}/\theta$ value beyond which $\chi^2-\chi^2_{\rm min}>9$.}
 \label{fig:chisq_vs_ratio}
\end{figure}

\begin{figure}[tbh]
    \centering
    \begin{subfigure}[b]{0.505\textwidth}
        \centering
        \includegraphics[width=\textwidth]{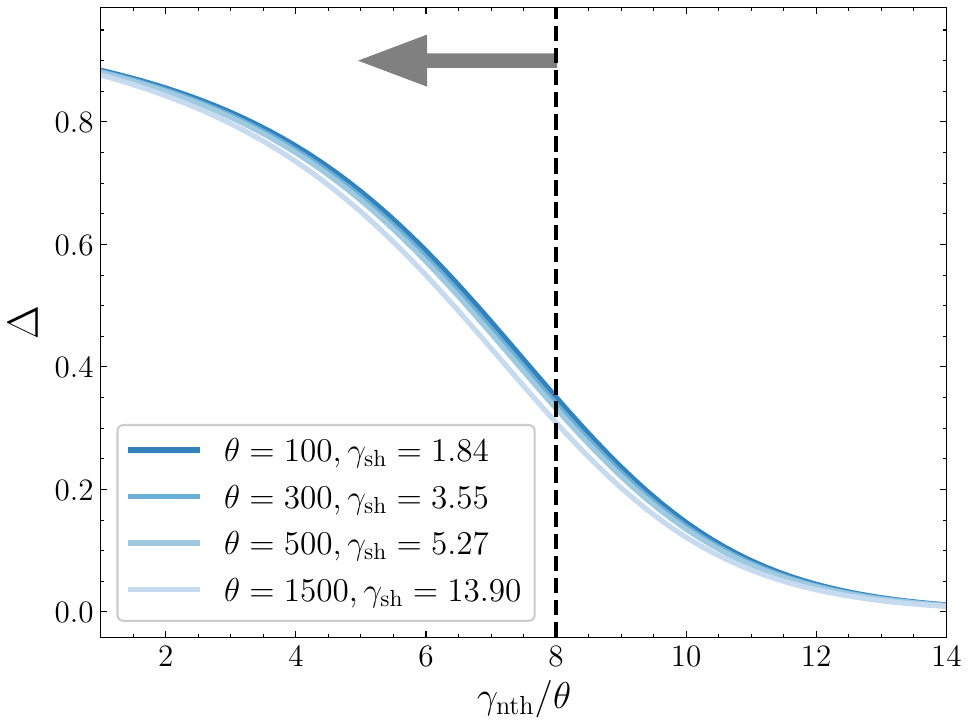}
    \end{subfigure}
    \begin{subfigure}[b]{0.52\textwidth}  
        \centering 
        \hspace{-0.2in}
        \includegraphics[width=\textwidth]{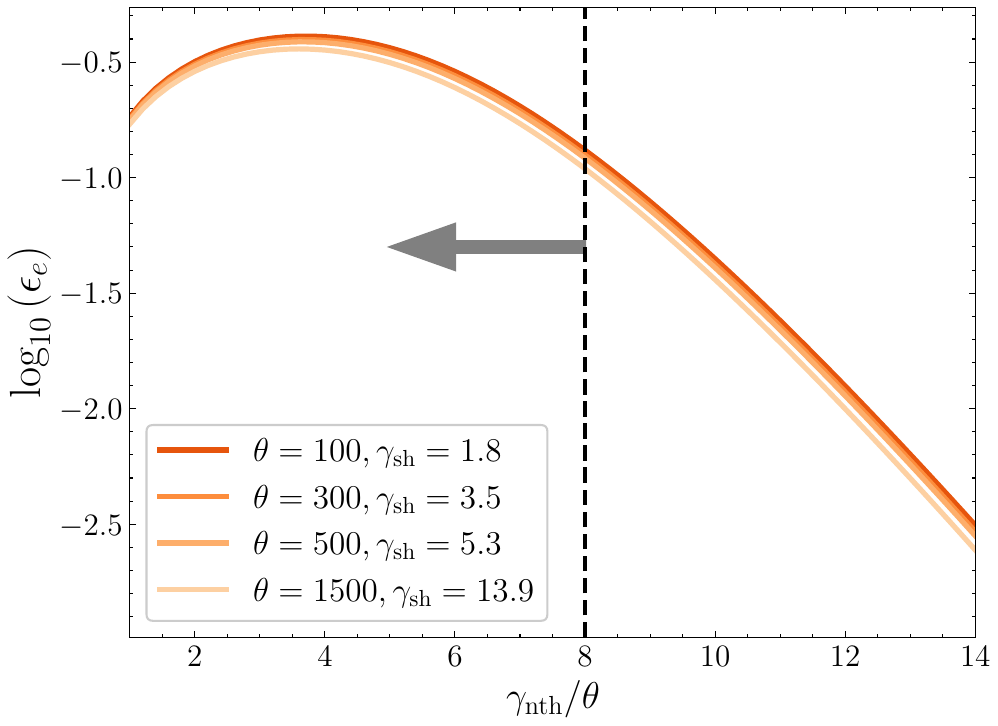}
    \end{subfigure}
    \caption{$\Delta$ (top panel) and $\epsilon_e$ (lower panel) as functions of $\gamma_{\rm nth}/\theta$ computed using Eq.~\ref{eq:Delta_def} and Eq.~\ref{eq:epsilon_e}, respectively. The allowed region lies left from the black vertical dashed line placed at $\gamma_{\rm nth}/\theta=8$, which is roughly the maximum value that we found compatible with the measurements.}
    \label{fig:epsilon_delta_constrain}
\end{figure}

\section{Results} \label{sec:results}

Figure~\ref{fig:chisq_vs_ratio} shows $\Delta\chi^2 = \chi^2-\chi^2_{\rm min}$ as a function of $\gamma_{\rm nth}/\theta$, where $\chi^2_{\rm min}$ is the minimum value for a given $\theta$. The scans were performed for $\theta=[100, 300, 500, 1500]$, corresponding to a shock velocity of $\gamma_{\rm sh}=[1.8, 3.5, 5.3, 13.9]$ (for $T_{\rm e}/T_{\rm p}=0.5$; see Sect.~\ref{sec:shock_simul}), representative of mildly-to-fully relativistic shocks (see below). The $\chi^2$ values increase significantly for high $\gamma_{\rm nth}/\theta \gtrsim 7-8$ due to an overestimation of the optical/UV and MeV-GeV fluxes when the thermal component becomes dominant. For $\theta=1500$, $\chi^2$ also increases in the range $\gamma_{\rm nth}/\theta \sim 2 - 6$. With such a large $\theta$, electrons emit photons primarily in the MeV-GeV band, and increasing $\gamma_{\rm nth}$ results in an underproduction of the \textit{Fermi}-LAT spectrum by the nonthermal electrons. For $\gamma_{\rm nth}/\theta \gtrsim 6$, the Maxwellian component becomes significant enough to compensate for this deficit. We emphasize that this deficit in the model could potentially be explained by invoking additional, unconstrained emitting regions in the jet. Consequently, Fig.~\ref{fig:chisq_vs_ratio} allows us to constrain only the upper limit of $\gamma_{\rm nth}/\theta$.\par

The $\chi^2$ values degrade beyond $\gamma_{\rm nth}/\theta \gtrsim 7 - 8$, irrespective of the value of $\theta$. Vertical dashed lines in Fig.~\ref{fig:chisq_vs_ratio} indicate, for each $\theta$, where $\chi^2-\chi^2_{\rm min} > 9$. We have chosen this threshold arbitrarily to roughly identify the range of $\gamma_{\rm nth}/\theta$ values that are inconsistent with the data, as this value is commonly used to set upper limits at the $3\sigma$ confidence level in $\chi^2$ tests.  We emphasize that a precise determination of the upper limits on $\gamma_{\rm nth}/\theta$ requires a more rigorous statistical analysis, including a proper consideration of the instrument response functions and instrumental systematics. This is beyond the scope of this work. In Appendix~\ref{sec:SED_scans}, we present the SED for each $\theta$ and with $\gamma_{\rm nth}/\theta$ in the transition region where the fits degrade, i.e. from $\gamma_{\rm nth}/\theta=6$ to $\gamma_{\rm nth}/\theta=10$. These SEDs clearly illustrate that, regardless of a detailed statistical treatment, $\gamma_{\rm nth}/\theta \gtrsim 8$ is inconsistent with the data.\par  

The model overproduces the radio emission at energies from $E\sim10^{-5}$\,eV to $E\sim10^{-3}$\,eV for $\theta < 500$. For $\theta \gtrsim 500$, the model exceeds the radio fluxes only when $\gamma_{\rm nth}/\theta \gtrsim 8$, but these values are already excluded by our parameter scan (see Fig.~\ref{fig:chisq_vs_ratio}). Consequently, our investigations simultaneously constrain $\gamma_{\rm nth}/\theta \lesssim 8$ \textit{and} $\theta \gtrsim 500$. Considering these constraints, and using the parameters in Table~\ref{tab:ssc_param}, our one-zone model achieves the best description of the optical-to-TeV data for $\theta \approx 500$ and $\gamma_{\rm nth}/\theta \approx 6$. The corresponding electron distribution is plotted with the darkest blue line in Fig.~\ref{eed_sample_distr}.\par 

To stay compatible with the upper limits from radio data even with $\theta \lesssim 500$, one may try to determine a parameter set providing an emitting region that is significantly more compact. In this case, the synchrotron self-absorption frequency may reach up to the millimeter range. However, as discussed in \citet{Tavecchio_Constraints}, the one-zone SSC model is a closed system. Once constraints from variability timescales and the (synchrotron) cooling break are included, there is very little freedom in choosing the parameter values. The observable $\nu_{\rm syn,peak}$ sets $\gamma_{\rm br}$, which is in turn dictated by the size of the emitting region and the magnetic field as a consequence of synchrotron cooling. Therefore, within our single-zone scenario, an agreement with the radio data for $\theta \lesssim 500$ is not possible. The only alternative is to make \textit{adhoc} assumptions on the shape of the electron distribution to include additional spectral breaks and/or to fully decouple $\gamma_{\rm br}$, $B$ and $R$ to allow a synchrotron self-absorption frequency in the millimeter range. In principle, one could also 
consider the possibility of a spectral break deviating from the standard cooling break of 1, perhaps due 
to inhomogeneities within the emitting region. These considerations, which have been discussed elsewhere \cite[e.g.,][]{abdo:2011,baheeja2022}, are however beyond the scope of the current study.\par

Figure~\ref{fig:epsilon_delta_constrain} gives the evolution of $\Delta$ and $\epsilon_e$ as function of $\gamma_{\rm nth}/\theta$. The region constrained by the SED of Mrk~421 lies left from the black vertical dashed line placed at $\gamma_{\rm nth}/\theta=8$, being roughly the maximum value that we found compatible with the measurements. The two quantities show little dependency on $\theta$. Therefore, independently of the choice of $\theta$, we infer that the particle distribution must be characterised by $\Delta \gtrsim 0.4$ and $\epsilon_e \gtrsim 10^{-1}$ in order to be in agreement with the data.\par 

To verify the energy requirements, we estimate the kinetic jet power as:
\begin{equation}
    L_{\rm jet, kin} \simeq \pi \; R^2 \; \Gamma_b^2 \; \beta c \; (u_{\rm p} + u_{\rm e} + u_{\rm B})  \ ,
\end{equation}
where $\Gamma_b \approx \delta$ is the bulk Lorentz factor of the emitting region in the observer's frame, $u_{\rm e} = n_{\rm e} \, \bar{\gamma} \, m_{\rm e} c^2$ the electron energy density, $u_{\rm B}=B^2/8\pi$ the magnetic energy density and $u_{\rm p} \approx n_{\rm p} \,m_{\rm p} c^2$ is the 
(cold) proton energy density. We assume an equal number density of protons and electrons, such that $n_{\rm p} = n_{\rm e}$. 
For all cases satisfying $\gamma_{\rm nth}/\theta \lesssim 8$, we find $L_{\rm jet, kin} \lesssim 10^{45}$\,
erg\,s$^{-1}$. This is well below the Eddington luminosity $L_{\rm Edd} \approx 3 \times 10^{46}$\,
erg\,s$^{-1}$ for the estimated black hole mass of Mrk~421 \citep[$M_{\rm BH} \simeq 2 \times 10^{8}$\,
M$_\odot$;][]{barth2003}. As expected, $L_{\rm jet, kin}$ decreases with increasing $\theta$ because the electron density is dominated 
by particles at the lowest energies. For our best-fit case ($\theta \approx 500$ \& $\gamma_{\rm nth}/\theta 
\approx 6$), the kinetic jet power is $L_{\rm jet, kin} \approx 5 \times 10^{43}$\,erg\,s$^{-1}$. 

\section{Relationship to shock parameters} \label{sec:shock_simul}

As demonstrated above, SED modelling in principle allows to place constraints on the shape of the thermal 
component. These constraints can, in turn, inform plasma simulations of mildly relativistic shocks. 
Conversely, simulations of shock physics can provide insight into the suitability of a given SED 
modelling. This requires, however, relating the inferred parameters to those commonly employed in 
simulations, which we address in the following.

We use $\gamma_0$ to denote the bulk Lorentz factor of the upstream plasma as measured in the downstream 
reference frame. At the shock, the incoming kinetic energy of the plasma is redistributed among thermal 
and nonthermal protons and electrons. For shocks that are accompanied by efficient particle acceleration, 
only a fraction $a_\mathrm{th} <1$ of the upstream energy is transferred to thermal protons and electrons,
i.e., we have: 
\begin{equation}
a_\mathrm{th}(\gamma_0-1)\,\mpr \, c^2=  (\gamma_\mathrm{th,e}-1)\,\me c^2 + 
(\gamma_\mathrm{th,p}-1)\, \mpr c^2,
\label{ene_redistr}
\end{equation}
where $\gamma_\mathrm{th,e}$ and $\gamma_\mathrm{th,p}$ are the average thermal electron and ion Lorentz 
factor at the shock downstream, respectively, and where we take $a_\mathrm{th} \approx0.8$ as fiducial 
reference value. Protons and electron are usually not in thermal equilibrium with each other downstream 
of the shock \citep{raymond23}, so that:
\begin{equation} 
\frac{\te}{\tp} = \frac{(\gamma_\mathrm{th,e}-1)\,\me c^2}{(\gamma_\mathrm{th,p}-1)\,\mpr c^2} < 1\ .
\label{tempratio}
\end{equation}
In PIC simulations of quasi-parallel, mildly relativistic ($\gamsh \simeq 2$) and 
weakly magnetized electron-ion shocks, for example, $T_e \sim 0.25\, T_p$ has been observed \citep{crumley2019}. 
Electron heating might be further enhanced for weakly magnetised \citep[$T_e \sim 0.3\, T_p$,][]{vanthiegem2020} 
or more relativistic shocks \citep[$T_e \sim 0.5\, T_p$,][]{sironi_spitkovsky_2011,sironi13}.\par 

Taking into account Eq.~\ref{ene_redistr}, Eq.~\ref{tempratio} and $\theta \equiv k T_{\rm e}/m_{\rm e} 
c^2 = \frac{1}{3} (\gamma_\mathrm{th,e}-1)$, the relationship between $\gamma_0$ and $\theta$ can be
expressed as:

\begin{equation}
\label{eq:gamma0totheta}
\gamz = 1 + \frac{3 \theta}{a_\mathrm{th}} \, \frac{\me}{ \mpr} \, \left(1+ \frac{\tp}{\te}\right)\,.
\end{equation}
The Lorentz factor of the shock $\gamsh$ and $\gamma_0$ are approximately related 
by~\citep{nishikawa09}:
\begin{equation}
\label{eq:gammashock}
\gamsh= \sqrt{\frac{(\gamz+1)(\gad(\gamz-1)+1)^2}{\gad(2-\gad)(\gamz-1)+2}} ,
\end{equation}
where $\gad=4/3$ is the adiabatic index of 3-dimensional relativistic gas. Taking
the best-fit case assuming a one-zone model, $\theta \approx 500$ (see Sect.~\ref{sec:results}), 
which yields $\gamma_{\rm sh} \approx 5.3$ (assuming $a_\mathrm{th} = 0.8$ and $\te/\tp=0.5$).
A key parameter in shock simulations is the magnetisation of the upstream plasma:
\begin{equation}
\label{eq:magnetization}
\sigma=\frac{B_0^2}{4 \pi \gamz n_0 (\mpr+\me) c^2} \ ,
\end{equation}
where $B_0$ is the upstream magnetic field and $n_0$ is the upstream plasma density measured in the 
downstream reference frame. SED modelling constraints the magnetic field $B$ and particle density $n$ 
of the downstream plasma measured in the comoving (downstream) reference frame. The upstream plasma 
density can be estimated using the shock compression ratio $r=n/n_0 = \frac{\gad(\gamz+1)}{\gamz(\gad-1)}$
\citep{nishikawa09}. Accordingly, the upstream plasma density becomes:
\begin{equation}
\label{eq:upstr_dens}
n_0=\frac{n\gamz(\gad-1)}{\gad(\gamz+1)} \ .
\end{equation}

The relationship between the magnetic field in the upstream and downstream regions of the shock 
is not straightforward. It depends on several factors, including magnetic field amplification in 
the foreshock region, compression at the shock, and decay in the downstream region. These processes, 
in turn, are influenced by the shock Lorentz factor, magnetisation, and obliquity. 
PIC simulations \citep[e.g.,][]{sironi_spitkovsky_2011,sironi13,crumley2019} indicate that 
approximately 10\% of the upstream kinetic energy in mildly relativistic shocks is converted into 
magnetic energy at the shock transition. However, these simulations also show that the self-generated
magnetic field decays rapidly in the downstream region, on scales of about a thousand of ion skin 
depths, $1000\lsi \sim 10^9$\,{cm}, which is much smaller than the size of the emitting region, 
$R \sim 10^{16}$\,cm. Since PIC simulations probe turbulence only on microscopic scales, it seems
possible that turbulence driven on larger scales may decay more slowly \cite[e.g.,][]{lemoine2013}, 
potentially influencing the entire emission region. Indeed, in the case of unmagnetized shocks 
\citep{chang08} mediated by the Weibel instability \citep{weibel}, turbulence on larger scales 
decays much slower. However, this topic remains largely unexplored, making a robust quantitative 
extrapolation to macroscopic scales difficult. In view of this, we assume the upstream magnetic 
field to be $B_0=B/a_\mathrm{B}$, where $a_\mathrm{B}$ represents the cumulative magnetic field 
amplification. In case of very weak magnetic field amplification or fast field decay, 
$a_\mathrm{B}=1$ corresponds to a parallel shock, while $a_\mathrm{B}=\gamz r$ corresponds to 
a perpendicular shock.

By combining the expressions for the plasma density and the magnetic field with 
Eq.~\ref{eq:magnetization}, the upstream plasma magnetisation can be expressed as:
\begin{equation}
\label{eq:magnetization_final}
\sigma=\frac{(B/a_\mathrm{B})^2}{4 \pi \gamz n (m_{\rm p}+\me) c^2} 
\frac{\gad(\gamz+1)} {[\gamz(\gad-1)]}\ ,
\end{equation}
In the case under consideration, with $\theta =500$, the electron density in the 
emitting region is $n\sim 1$\,cm$^{-3}$ and the upstream plasma magnetisation is 
constrained to $\sigma \sim 0.15/a_{\rm B}^2$.

\section{Conclusions} \label{sec:discussion}

PIC simulations of plasma shocks demonstrate the development of a thermal component in the low-energy 
part of the distribution \citep[e.g.,][]{sironi_spitkovsky_2011, crumley2019}. For electron-ion shocks, 
the electron thermal component can easily peak at Lorentz factors as high as $\gamma \sim 10^2 - 10^3$,
effectively mimicking an electron distribution with high $\gamma_{\rm min}$. In this work, we have modelled 
the broadband emission of the BL Lac type object Mrk~421 by adopting an electron distribution composed of 
a (relativistic) thermal Maxwell distribution and a nonthermal (broken) power-law component. 
While observations (from radio frequencies to TeV) do not allow us to claim the existence of a thermal 
(Maxwellian) population, such a component is expected from first principles. In this work, we have shown 
that the broadband data greatly restrict the contribution of the Maxwellian electrons, providing 
constraints on the physical properties of the shock.\par

To avoid overproducing the observed fluxes in the optical/UV and MeV-GeV bands (where the thermal 
electrons emit most of their radiation), we find that the ratio between the Lorentz factor at which 
the nonthermal distribution begins ($\gamma_{\rm nth}$) and the dimensionless electron temperature
($\theta$) must satisfy $\gamma_{\rm nth}/\theta \lesssim 8$. 
Since $\gamma_{\rm nth}/\theta$ controls the fraction $\Delta$ of the total electron energy residing 
in the nonthermal component, and the fraction $\epsilon_e$ of the total shock energy carried by the 
nonthermal electrons, the limit $\gamma_{\rm nth}/\theta \lesssim 8$ translates into the following 
constraints: $\Delta \gtrsim 0.4$ and $\epsilon_e\gtrsim 10^{-1}$. Such a limit for $\epsilon_e$ 
seems broadly consistent with simulations of relativistic shock \citep[e.g.,][]{sironi_spitkovsky_2011}.
Dedicated PIC simulations will be needed to verify the extent to which these results can be reproduced 
under the inferred conditions, a task we intend to address in a subsequent study. It is important to 
note that the constraints we obtain are only slightly dependent on the value of $\theta$, and thus 
the shock velocity (see Sect.~\ref{sec:shock_simul}). However, to stay below the radio fluxes, 
$\theta \gtrsim 500$, which suggests that $\gamma_{\rm sh} \gtrsim 5.3$ assuming $a_\mathrm{th} = 0.8$ 
and $\te/\tp=0.5$ and using Eq.~\ref{eq:gammashock}. Our analysis therefore favours shocks in the mildly 
or fully relativistic regime.\par 

Broadband modelling of HSPs, ``extreme TeV'' blazars and and recent X-ray polarisation results suggest
a relatively high $\gamma_{\rm min}$, up to $\sim 10^4$ \citep[see e.g.,][]{costamante2018, magic_ixpe_mrk421flare2023}. 
While the Maxwellian component is often neglected in the literature, if one were to equate $\gamma_{\rm min}$ with $\gamma_{\rm nth}$, then $\gamma_{\rm min} \sim 10^4$ would imply $\theta\sim 1000$ and $\gamma_{\rm nth}/\theta \sim 8$, a scenario still consistent with our parameter scan. Alternatively, as discussed in \citet{zech_lemoine2021}, a sequence of multiple shocks could elevate an initial (lower) value of $\gamma_{\rm min}$, potentially leading to the values considered here.\par

The model employed here considers a single emission zone, whereas multiple regions within the jet may contribute to the overall SED. This limitation explains why we can only establish upper limits on $\gamma_{\rm nth}/\theta$, and consequently, our conclusions remain largely unaffected by the potential contribution of multiple emission zones. However, the interaction of electrons with the radiation field from other jet regions \citep[as in the ``spine-layer'' model;][]{ghisellini2005} could alter the inverse-Compton spectral shape compared to that predicted by the single-zone model. This alteration may influence our limits, as they are directly dependent on the radiative signatures observed in the gamma-ray band. A detailed study is necessary to accurately evaluate this impact. Nevertheless, we anticipate that achieving a satisfactory description of the gamma-ray data for $\gamma_{\rm nth}/\theta > 8$ would necessitate a significant (and unnatural) fine-tuning of the external photon field spectrum, given the distinct (and broad) signature of the Maxwellian electrons (see Appendix~\ref{sec:SED_scans}).\par

Under the assumption that a single zone is responsible for the emission from optical to TeV, our best fit 
is obtained for $\gamma_{\rm nth}/\theta \approx 6$ and $\theta \approx 500$, i.e., 
$\gamma_{\rm sh} \approx 5.3$ (again for $a_\mathrm{th} = 0.8$ and $\te/\tp=0.5$). This could be compared
to simulations of subluminal mildly-relativistic shocks where strong electron heating and efficient 
particle acceleration with $\epsilon_{\rm e}\approx 0.1$ has been observed \citep{sironi_spitkovsky_2011}. 
In the framework of a 'blob-in-jet' scenario, where a fast ($\Gamma_b$) obstacle overtakes the jet 
($\Gamma_j < \Gamma_b$), $\gamma_{\rm sh} \simeq \Gamma_b/2\Gamma_j$ \cite[e.g.,][]{zech_lemoine2021}. 
For a closely aligned jet, where $\delta \sim \Gamma_b$, inferred mean jet flow speeds would then be 
of the order $\Gamma_j \sim \Gamma_b/2\gamma_{\rm sh} \sim$ a few. Considering a stationary shock, the jet Lorentz factor must obey $\Gamma_j \gtrsim \gamma_{\rm sh} \ \delta /2 \sim 50$. Such a value is at the extreme of the distribution of Lorentz factors obtained from VLBI observations of parsec-scale jets \citep{Homan2021}.\par

The temperature ratio $T_e/T_p=0.5$ assumed throughout this work is compatible with PIC simulations of relativistic electron-ion shocks, as reported in \citet{sironi_spitkovsky_2011}. Similar values are found by \citet{crumley2019} for the mildly-relativistic case. We note that the temperature ratio is sensitive to other other parameters such as the magnetisation. Nevertheless, the results reported in the works mentioned above cover a range of magnetisation (and $\gamma_{\rm sh}$) that is comparable to the one derived using our phenomenological model. Therefore, assuming $T_e/T_p=0.5$ is a well-motivated initial guess, and our conclusions also do not strongly depend on the exact value of $T_e/T_p$. To obtain a more accurate determination of the temperature ratio, one would need to explore a wide range of $\gamma_{0}$ and magnetization with PIC simulations, which we leave for a future work.\par

Our model was applied to the average emission state of Mrk~421 observed in 2009, a period characterized by an extended phase of low flux variability \citep{abdo:2011}. Rapid outbursts, frequently observed in blazars and, by definition, representing peculiar source states, are potentially caused by shocks with distinct physical conditions compared to those associated with the quasi-stable, quiescent emission. Further investigation into the connection between shock dynamics and emission states will be crucial for a complete understanding of the mechanisms driving rapid outbursts. We defer the investigation of flaring events to future work.\par

The results presented here contribute to the development of a comprehensive model for understanding shock acceleration in blazar jets. High-quality SED characterisations, such as the ones that can be achieved with long multi-instrument observations of bright blazars, combined with dedicated simulations, will play a key role in further advancements in this model.

\begin{acknowledgements}
We thank Lea Heckmann for fruitful discussions in the early phases of the project, as well as Martin Lemoine for useful exchanges. We also acknowledge support from the Deutsche Forschungs gemeinschaft (DFG, German Research Foundation) under Germany’s Excellence Strategy – EXC-2094 – 390783311.

\end{acknowledgements}

\bibliographystyle{aa}
\bibliography{bibliography_paper.bib}

\begin{appendix}
\section{SED scans}
\label{sec:SED_scans}

\begin{figure*}[tbh]
    \centering
    \begin{subfigure}[b]{0.49\textwidth}
        \centering
        \includegraphics[width=\textwidth]{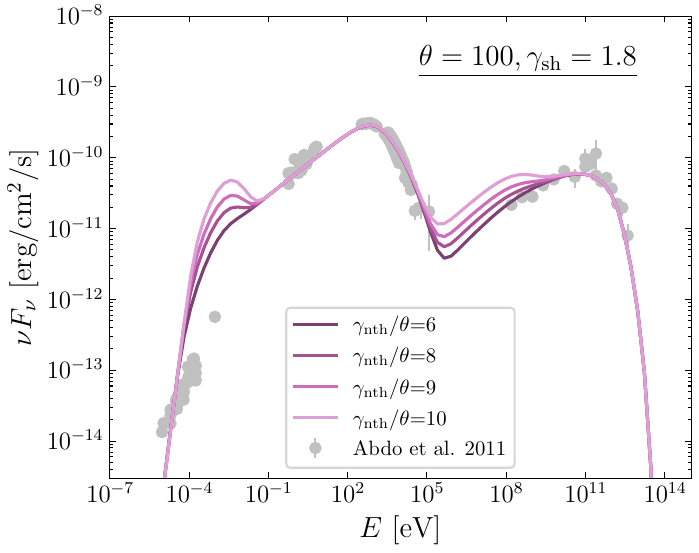}
    \end{subfigure}
    \begin{subfigure}[b]{0.49\textwidth}  
        \centering 
        \includegraphics[width=\textwidth]{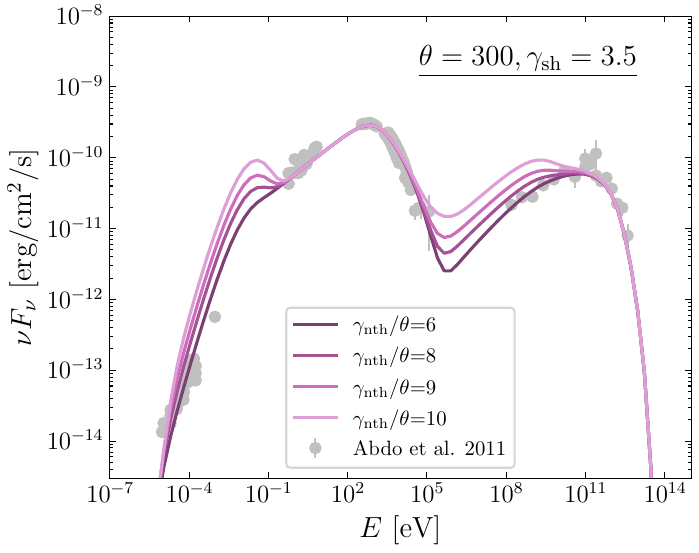}
    \end{subfigure}
        \centering
    \begin{subfigure}[b]{0.49\textwidth}
        \centering
        \includegraphics[width=\textwidth]{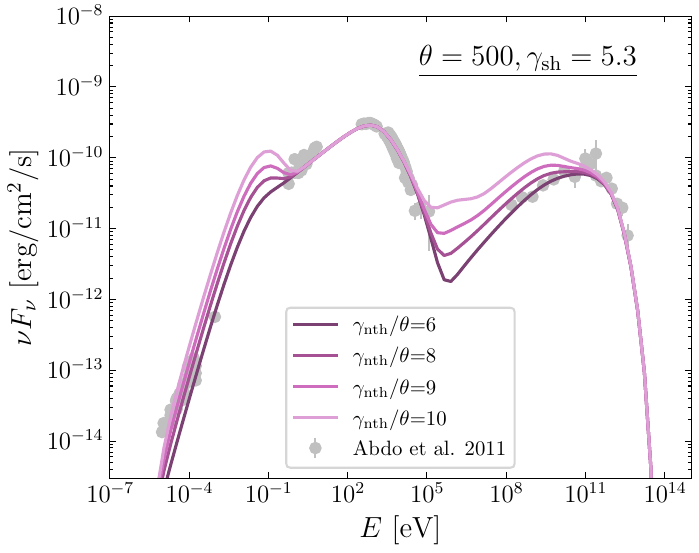}
    \end{subfigure}
    \begin{subfigure}[b]{0.49\textwidth}  
        \centering 
        \includegraphics[width=\textwidth]{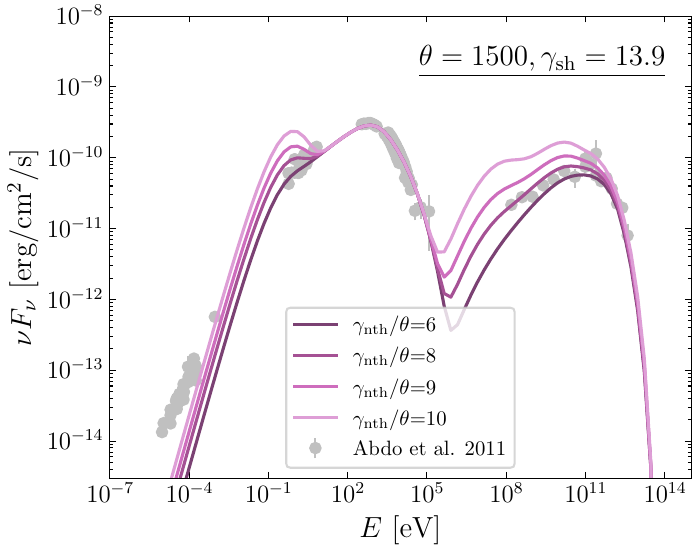}
    \end{subfigure}
    \caption{SED models for $\theta=[100,300,500, 1500]$. Solid lines, from light violet to dark violet, cover three different $\gamma_{\rm nth}/\theta = [6, 8, 9, 9]$, i.e. the transition region where the fit degrades. For each $\theta$, the corresponding shock velocity $\gamma_{\rm sh}$ is also given. We refer the reader to Sect.~\ref{sec:shock_simul} for more details on the connection between $\theta$ and $\gamma_{\rm sh}$. The SED from \cite{abdo:2011} is plotted with grey markers.}

    \label{fig:scan_SED}
    
\end{figure*}

Figure~\ref{fig:scan_SED} presents the SED model for all values of $\theta$ considered in this study, $\theta=[100, 300, 500, 1500]$. The model is shown for a few values of $\gamma_{\rm nth}/\theta$ around the transition region where the fit significantly degrades. 

\end{appendix}

\end{document}